\documentclass[twocolumn,aps]{revtex4}
\usepackage{graphicx}
\usepackage{amsmath}

\begin{document}

\title{Experimental nonlinear sign shift for linear optics quantum computation}

\author{G. Kaoru Sanaka$^1$, Thomas Jennewein$^1$,  Jian-Wei Pan$^1$, Kevin Resch$^1$,, and Anton
Zeilinger$^1,^2$}

\affiliation{$^1$ Institut f$\ddot{u}$r Experimentalphysik,
Universit$\ddot{a}$t Wien, Boltzmanngasse 5, A-1090 Wien,
Austria\\ $^2$ Institut f$\ddot{u}$r Quantenoptik und
Quanteninformation, $\ddot{O}$sterreichische Akademie der
Wissenschaften, Boltzmanngasse 3, A-1090 Wien, Austria}

\begin{abstract}
We have realized the nonlinear sign shift (NS) operation for
photonic qubits.This operation shifts the phase of two photons
reflected by a beam splitter using an extra single photon and
measurement. We show that the conditional phase shift is $\left(
1.05\pm 0.06\right) \pi $ in clear agreement with theory. Our
results show that by using an ancilla photon and conditional
detection, nonlinear optical effects can be implemented using
only linear optical elements. This experiment represents an
essential step for linear optical implementations of scalable
quantum computation.
\end{abstract}

\date{\today}

\maketitle

A promising system for quantum computation is to use single
photons to encode quantum information \cite{Chuang,Milburn}. This
is due to the photon's robustness against decoherence and the
availability of single-qubit operations. However it has been very
difficult to achieve the necessary two-qubit operations since the
physical interaction between photons is much too small.
Surprisingly, Knill, Laflamme, and Milburn (KLM) showed that
effective nonlinear interactions can be implemented using only
linear optical elements in such a way that scalable quantum
computation can be achieved \cite
{Knill,Gottesman,Ralph,Koashi,Pittman,Zou,Hofmann,Sanaka}. The
fundamental element of the KLM scheme is the nonlinear sign-shift
(NS) operation from which the two-qubit conditional sign flip
gate can be constructed. Universal quantum computation is then
possible with this two-qubit gate together with all single-qubit
rotations \cite{Sleator,Barenco} Here, we experimentally
demonstrate the NS operation using photons produced via parametric
down-conversion. In contrast with the KLM scheme, our method to
observe the NS operates in the polarization basis and therefore
does not require interferometric phase stability.

\begin{figure}[tbp]
\includegraphics{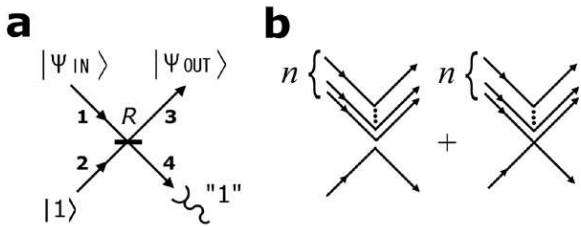}
\caption{(a) Schematic of a simplified version of nonlinear
sign-shift (NS) operation constructed by a non-polarizing beam
splitter of reflectivity $R$. $|\Psi _{\text{IN}}\rangle $ and
$|\Psi _{\text{OUT}}\rangle $ are the quantum states of input and
output photons. The operation is successful when the
single-photon detector in ancilla mode 4 counts a single photon.
(b) The two paths that lead to the detection of exactly one
photon in output mode 4. As long as all of the photons are
indistinguishable, these two paths can interfere.} \label{fig1}
\end{figure}

A simplified version of NS operation is shown in Fig.\
\ref{fig1}(a). An input state, $|\Psi _{\text{IN}}\rangle
=|n\rangle $, impinges on a beam-splitter (BS) with reflection
probability $R$; a single ancilla photon, $|1\rangle \,$impinges
from the other side of the beam splitter. The two input modes, 1
and 2, undergo a unitary transformation into two output
modes, 3 and 4, described by $\widehat{a}_{1}\rightarrow \sqrt{R}\,\widehat{a%
}_{3}+\sqrt{1-R}\,\widehat{a}_{4}$ and $\widehat{a}_{2}\rightarrow -\sqrt{1-R%
}\,\widehat{a}_{3}+\sqrt{R}\,\widehat{a}_{4}\,$. The NS operation
is successful when one and only one photon reaches the detector
in mode 4. \ Provided the photons are indistinguishable, the two
paths leading to exactly one photon in mode 4 will interfere. The
two interfering processes are depicted in Fig.\ \ref{fig1}(b) for
$n$ input photons. Either all $n+1$ photons are reflected or
$n-1$ of the photons in mode 1 are reflected and 1 photon in each
of modes 1 and 2 are transmitted. When a single photon ends up in
mode 4, the photon number state undergoes the following
transformation:

\begin{equation}
|\Psi _{\text{IN}}\rangle =|n\rangle \rightarrow |\Psi
_{\text{OUT}}\rangle =(\sqrt{R})^{n-1}[R-n\,(1-R)]\,|n\rangle ,
\label{equ1}
\end{equation}
where the unusual normalization of the output state reflects the
probability amplitude of success. The sign of the phase shift
depends on the number of incident photons and the reflection
probability of the BS. For $n<R/(1-R)$, the sign of the amplitude
is unchanged and for $n>R/(1-R)$ it picks up a negative sign. For
the critical case, where $n=R/(1-R)$, the output probability
amplitude becomes zero \cite{Hong}.

In the original KLM proposal, the NS gate is achieved using a
phase sensitive interferometer. \ In our experiment, we induce
the phase shift between two polarizations in the same spatial
mode and therefore have much less stringent stability
requirements. \ The extension of the NS operation to include a
second polarization mode is straightforward. We inject a
horizontally-polarized ancilla photon into the BS in Fig.\
\ref{fig1} (a) and consider only the cases where the single
photon detected in mode 4 is horizontally polarized. The
transformation for the horizontal polarization is the same as in
Eq.(\ref{equ1}). There is only 1 possible path which leads to no
vertically-polarized photons in mode 4; that is for all
vertically-polarized photons to be reflected. This operation for
the input state with $m$ vertically-polarized photons and $n$
horizontally-polarized photons is given by:
\begin{eqnarray}
&& |\Psi _{\text{IN}}\rangle =|m_{\text{V}};n_{\text{H}}\rangle
\stackrel{\text{NS}}{\rightarrow }|\Psi _{\text{OUT}}\rangle =
 \nonumber \\ && (\sqrt{R_{%
\text{V}}})^{m}(\sqrt{R_{\text{H}}})^{n-1}[R_{\text{H}}-n_{\text{H}}(1-R_{%
\text{H}})]\,|m_{\text{V}};n_{\text{H}}\rangle ,  \label{equ2}
\end{eqnarray}
where $R_{\text{V}}$ and $R_{\text{H}}$ are the reflection
probabilities for vertical and horizontal polarization
respectively. As expected, the vertical photon number, $m$, does
not appear in the square bracket nonlinear-sign term. The only
change the vertically-polarized photons contribute is the
reflection amplitude raised to the power of $m$.

A quantum phase gate for the KLM scheme can be implemented using two such NS gates when the BS has reflection probabilities, $R_{\text{V}}=5-3\sqrt{2} \approx 0.76$ and $R_{\text{H}}=(3-\sqrt{2})/7 \approx 0.23$ \cite{Sanaka2}. For the experiment we use input states where $m+n=2$ and the typical 50/50 BS, where $R_{\text{V}}=R_{\text{H}}=1/2$. The three possible input states are transformed by the NS operation according to:{%
\setcounter{enumi}{\value{equation}} \addtocounter{enumi}{1} %
\setcounter{equation}{0} \renewcommand{\theequation}{\theenumi%
\alph{equation}}
\begin{eqnarray}
&&|2_{\text{V}};0_{\text{H}}\rangle \rightarrow \frac{1}{2\sqrt{2}}\,|2_{%
\text{V}};0_{\text{H}}\rangle ,  \label{equ3a} \\
&&|1_{\text{V}};1_{\text{H}}\rangle \rightarrow 0\,,  \label{equ3b} \\
&&|0_{\text{V}};2_{\text{H}}\rangle \rightarrow -\frac{1}{2\sqrt{2}}\,|0_{%
\text{V}};2_{\text{H}}\rangle ,  \label{equ3c}
\end{eqnarray}
\setcounter{equation}{\value{enumi}} }\newline The operation with
this set of input parameters serves to change only the phase of
the input state $|0_{\text{V}};2_{\text{H}}\rangle $. The input
state $|1_{\text{V}};1_{\text{H}}\rangle $ is ``annihilated'' by
this operation \cite{Hong}. This means that for that input state
the condition of having exactly one horizontally polarized photon
in mode 4 never occurs. The NS operation using this particular BS
reflectivity is important to a related protocol for a ``quantum
filter'' \cite{Hofmann2}.

\begin{figure}[tbp]
\includegraphics{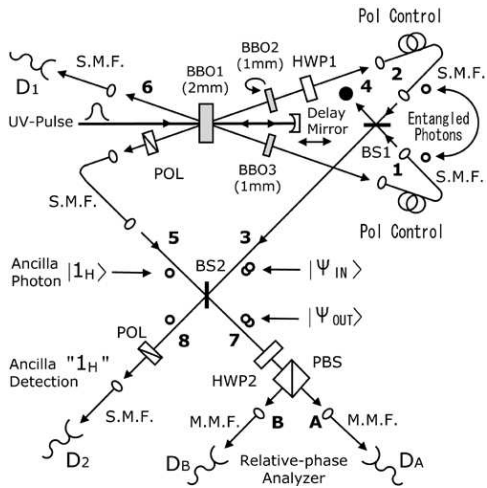}
\caption{Experimental setup for the demonstration of nonlinear
sign-shift (NS) operation using double-pass parametric
down-conversion. Photon pairs created from first pass are used
for the input of NS operation and pairs from the second pass are
used for the triggered single photon source as ancilla.
Successful operation is identified through four-fold coincidence
counts between all four detectors.} \label{fig2}
\end{figure}

In the experiment (Fig.\ \ref{fig2}), frequency-doubled pulses
from a mode-locked Ti:Sapphire laser (center wavelength 394.5nm,
200 fs pulse duration, 76 MHz repetition rate) make two passes
through a type-II phase-matched 2-mm BBO crystal (BBO1)
\cite{Bouwmeester,Pan,Jennewein,Pan2}. Through spontaneous
parametric down-conversion there is some probability for one pair
of entangled photons to be created on the first pass and another
pair on the second \cite{Kwiat2}. Additional 1-mm crystals (BBO2,
BBO3) and a half-wave plate (HWP1) are used for the compensation
of the birefringence effect inside BBO1 and also for the
selection of appropriate polarization-entangled photons. The
first pair (right going modes 1 and 2 in Fig.\ \ref{fig2}) serves
as the input to the NS operation in mode 3. The second pair (left
going modes 5 and 6) is used to produce the ancilla photon. Upon
detection of a ``trigger'' photon in mode 6, a single photon
state will be present in mode 5 with high probability. BS2 is a
normal 50/50 BS and its reflectivity determines the NS operation.
Four-photon events are
post-selected by single-photon counting detectors D$_{1}$, D$_{2}$, D$_{%
\text{A}}$, and D$_{\text{B}}$. We first verified the operations (\ref{equ3a}%
) and (\ref{equ3c}). The state of the polarization-entangled
photons in mode
1 and 2 was prepared as $|\Phi _{\theta }\rangle =1/\sqrt{2}(|1_{\text{V}%
}\rangle _{1}|1_{\text{V}}\rangle _{2}+e^{i\theta
}|1_{\text{H}}\rangle _{1}|1_{\text{H}}\rangle _{2}\,)$, where
the relative phase, $\theta $, was controlled by tilting the
compensation crystal BBO2. BS1 (also 50/50) transforms this
polarization-entangled state to a photon-number entangled state.
The photons in mode 3 is in the state $|\Psi _{\text{IN}}\rangle
=1/\sqrt{2}(|2_{\text{V}};0_{\text{H}}\rangle _{3}+e^{i\theta
}\,|0_{\text{V}};2_{\text{H}}\rangle _{3})$. Pairs created from
the second pass are used as a source of triggered single photons.
A translatable mirror on the pump allows for the relative
creation time of the two pairs to be varied. Down-converted
photons are coupled into single-mode fibers (S.M.F.) for mode
filtering. The photons come out of the fibers to free space again
for interference in BS2 (also 50/50 BS). The polarization of the
ancilla photon $|1_{\text{H}}\rangle $ in mode 5 is set to
horizontal using a polarizer (POL). The entangled photons are
sent to BS2 and combined with the horizontally polarized ancilla
photon in mode 5. A successful operation occurs when the single
photon detector, D$_{2}$, in output mode 8 counts a single photon
in horizontal polarization state. \ The output state in mode 7 is
analyzed using HWP2, a polarizing beam splitter (PBS), multi-mode
fibers (M.M.F.) and single-photon counters placed in modes A and
B (D$_{\text{A}}$and D$_{\text{B}}$) -- together these form a
relative-phase analyzer. Successful operation is identified
through the four-fold
coincidence counts between all four detectors (D$_{1}$, D$_{2}$, D$_{\text{A}%
}\,$and D$_{\text{B}}$ ).

In general, D$_{2}$, would need to be able to distinguish one
photon from multiple photons. However, in a multi-photon
coincidence experiment events with five or more photons are of
much lower probability and contribute negligibly to the signal.
The production probably of down-converted photons to the
measurement is estimated to be about $10^{-11}$ with four-photon
events (two-photon pair) per UV pulse of pump beam, and less than
$10^{-16}$ with five or more photon events (more than
three-photon pair). This post-selection process allows for the NS
operation signal to be observed using current generation
photodetectors and probabilistic signal photon sources.
Eq.(\ref{equ3a}) and Eq.(\ref{equ3c}) show that the maximum
successful probability of the NS operation is 1/8, however the
experimental results become much lower (about $10^{-5}$) because
we use pulsed down-converted photons for the operation instead of
ideal single-photon sources. Further developments of novel
single-photon sources and photo detectors will allow for
subsequent refinements of this experiment, such as eliminating
the need for post selection and pushing the experimental
probability of success towards its theoretical maximum
\cite{Kurtsiefer,Michler,Lounis,Santori,Kim,Takeuchi}.

\begin{figure}[tbp]
\includegraphics{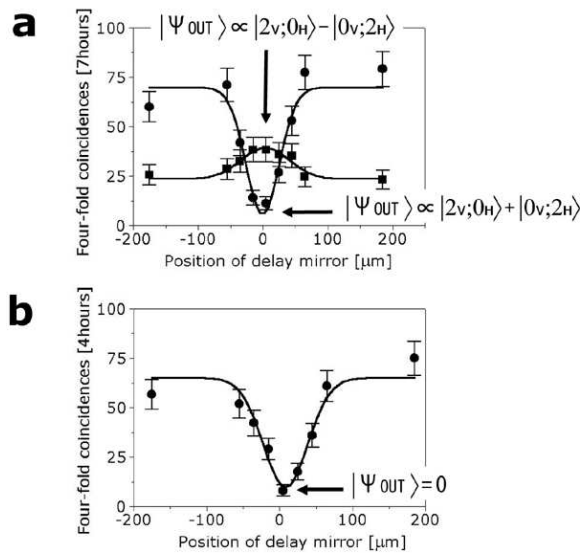}
\caption{(a) The four-fold coincidences as a function of the pump
delay mirror position for the photon-number entangled photons.
HWP2 is set to rotate the polarization state by 45 degrees. Solid
squares (circles) show the four-fold coincidence counts for the
input state where $\theta =0 \ (\pi)$. The peak and dip of the
two curves are different in size only because of accidental
coincidences occurring when the delay is much larger than the
photons' coherence length. (b) The four-fold coincidences as
functions of the pump
delay mirror position for the input photons in the state $|1_{\text{V}};1_{%
\text{H}}\rangle _{3}$. The HWP2 is set to leave the input
polarization states unchanged. At zero delay, the coincidences
are suppressed nearly to zero by the HOM effect.} \label{fig3}
\end{figure}

For our first measurement we act on the photon-number entangled
states in mode 3. Since the NS operation is an interference
effect, it only proceeds when the entangled photons and the
ancilla photon arrive at BS2 within their coherence time, $\tau
_{coh}$. In this case, the operation performs the following:
\begin{eqnarray}
|\Psi _{\text{IN}}\rangle &=&\frac{1}{\sqrt{2}}(|2_{\text{V}};0_{\text{H}%
}\rangle _{3}+e^{i\theta }\,|0_{\text{V}};2_{\text{H}}\rangle
_{3})
\nonumber \\
&\rightarrow &|\Psi _{\text{OUT}}\rangle =\frac{1}{4}(|2_{\text{V}};0_{\text{%
H}}\rangle _{7}-e^{i\theta }\,|0_{\text{V}};2_{\text{H}}\rangle
_{7}), \label{equ4}
\end{eqnarray}
To analyze this effect, we use HWP2 to rotate the polarization states by 45$%
^{\circ }$. Under such an operation, $1/\sqrt{2}(|2_{\text{V}};0_{\text{H}%
}\rangle +\,|0_{\text{V}};2_{\text{H}}\rangle )\rightarrow 1/\sqrt{2}(|2_{%
\text{V}};0_{\text{H}}\rangle
+\,|0_{\text{V}};2_{\text{H}}\rangle )$ (i.e.
it is invariant), however $1/\sqrt{2}(|2_{\text{V}};0_{\text{H}}\rangle -|0_{%
\text{V}};2_{\text{H}}\rangle )\rightarrow
|1_{\text{V}};1_{\text{H}}\rangle $ (i.e. it can produce
coincidences between D$_{\text{A}} $ and D$_{\text{B}}$).  We can
verify the transformation (\ref{equ4}) by measuring one
vertically- and one horizontally-polarized photon using the PBS
and photo detectors D$_{\text{A}}$ and D$_{\text{B}}$. Fig.\
\ref{fig3}(a) shows the observed variations of the count rate as
functions of the pump mirror position which varies the relative
arrival times of the entangled photons and the ancilla photons at
BS2. Solid squares (circles) show the four-fold coincidence
counts for $\theta =0 \ (\pi)$ in (\ref{equ4}) as a function of
the pump delay.  At zero delay, the coincidences for
D$_{1}$D$_{2}$D$_{\text{A}}$D$_{\text{B}}$ are enhanced
(suppressed) by the NS operation (\ref{equ4}). When the arrival
time difference is larger than $\tau _{coh}$, we obtain
coincidence counts from the state $|\Psi _{\text{IN}}\rangle$ in
(\ref{equ4}) and also accidental coincidence counts between the
ancilla photon and one entangled photon. If one could resolve the
sub-picosecond level time differences, the size of the dip of one
curve would be equal to the size of the peak in the other. The
fidelity of the sign-shifted entangled photons can be estimated
from the data to be $77\pm 6\%$. When taking the fidelity
$92.9\pm 0.5\%$ of the initial state into account this confirms
the high quality of our NS operation.

\begin{figure}[tbp]
\includegraphics{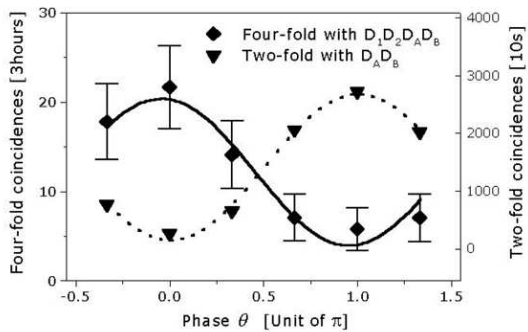}
\caption{The observed variation of the coincidence count rates as
functions of the phase of entangled photons $\theta $ at zero
delay for the photon-number entangled photons. Solid triangles
represent the two-fold coincidences with
D$_{\text{A}}$D$_{\text{B}}$ and show the phase of input
photons. Solid diamonds represent the four-fold coincidences with D$_{1}$D$%
_{2}$D$_{\text{A}}$D$_{\text{B}}$ which show the phase of the
output photons demonstrating a successful NS operation. Error
bars are based on the usual Poisson fluctuation in the number of
counts on the uncorrected data (Error bars of two-fold
coincidences are too small to display). The phase of four-fold
coincidence is shifted $(1.05\pm 0.06)\pi $ against two-fold
coincidence in agreement with the expected $\pi $ phase shift.}
\label{fig4}
\end{figure}

We then fix the pump delay at zero so that the NS operation can
proceed with maximum efficiency. We analyze the output mode 7 by
tilting the compensation
crystal BBO2 - this allows for the variation of the correlations with $%
\theta $ to be directly observed. For the state $|2_{\text{V}};0_{\text{H}%
}\rangle +e^{i\theta }\,|0_{\text{V}};2_{\text{H}}\rangle $ in
mode 7, the coincidence rate between detectors D$_{\text{A}}$ and
D$_{\text{B}}$ is
proportional to $\sin ^{2}(\theta /2)$. We input the state $|2_{\text{V}};0_{%
\text{H}}\rangle +e^{i\theta }\,|0_{\text{V}};2_{\text{H}}\rangle
$ into BS2
and record both the two-fold D$_{\text{A}}$D$_{\text{B}}$ and the four-fold D%
$_{1}$D$_{2}$D$_{\text{A}}$D$_{\text{B}}$ coincidences with the
ancilla path open. The two-fold coincidence rate reflects the
initial correlations for the input entangled-photon pair, whereas
the four-fold coincidence rate shows the correlations after a
successful NS operation. Fig.\ \ref{fig4} shows the observed
coincidence rates (two-fold are solid triangles and four-fold are
solid diamonds) at zero delay for different phase angles set by
BBO2. Clearly the phase of the correlations has been changed by
the NS operation; the relative phase between the two curves is
$(1.05\pm 0.06)\pi $ in excellent agreement with the expected
shift of $\pi $.

To complete the experimental confirmation of the NS operation we
also verified its action on the input state
$|1_{\text{V}};1_{\text{H}}\rangle $
(Eq.\ref{equ3b}). We prepare the input state $|\Psi _{\text{IN}}\rangle =|1_{%
\text{V}};1_{\text{H}}\rangle _{3}$ from the
polarization-entangled photons that was prepared as $|\Psi
^{+}\rangle =1/\sqrt{2}(|1_{\text{V}}\rangle
_{1}|1_{\text{H}}\rangle _{2}+|1_{\text{H}}\rangle
_{1}|1_{\text{V}}\rangle _{2}\,)$. The HWP2 is set such that it
does not rotate the polarization. Fig.\ \ref{fig3}(b) shows the
four-fold coincidences as a function of the
pump delay. At zero delay, the four-fold coincidences D$_{1}$D$_{2}$D$_{%
\text{A}}$D$_{\text{B}}$ are suppressed nearly to zero because of
the effect at BS2\cite{Hong}. The visibility of the fringe is
about $89\pm 4\%$. The results shown in Fig.\ \ref{fig3} and
Fig.\ \ref{fig4} confirm all of the important features of the NS
operation for input states with 2 photons as described
theoretically in Eq.(\ref{equ3a} - \ref{equ3c}).

We have experimentally demonstrated the nonlinear sign shift
operation using linear optical elements and the best available
technologies for single-photon generation and detection. This
includes using a triggered single-photon source from parametric
down-conversion and single-photon counting by avalanche
photodiodes. This experiment is a proof-in-principle demonstration
of the operation of the nonlinear sign shift, which is the
critical element in the KLM scheme. These experimental results
are of utmost importance for the realization of scalable quantum
computer with linear optics.

This work was supported by the Austrian Science Foundation (FWF),
project numbers M666 and SFB 015 P06, NSERC, and the European
Commission, contract number IST-2001-38864.

\end{document}